\documentclass{article}
\usepackage[preprint]{spconf}
\usepackage{amsmath,graphicx}

\title{Deep Performer: Score-to-Audio Music Performance Synthesis}

\name{Hao-Wen Dong~\textsuperscript{1,2~\sthanks{Work done during an internship at Dolby. Contact: \href{mailto:hwdong@ucsd.edu}{\texttt{hwdong@ucsd.edu}}}} \qquad Cong Zhou~\textsuperscript{1} \qquad Taylor Berg-Kirkpatrick~\textsuperscript{2} \qquad Julian McAuley~\textsuperscript{2}}

\address{\textsuperscript{1}~Dolby Laboratories \qquad \textsuperscript{2}~University of California San Diego}

\usepackage[hidelinks]{hyperref}
\usepackage{url}
\usepackage{booktabs}
\usepackage{cite}
\usepackage{enumitem}
\usepackage{bm}
\usepackage{amssymb}
\usepackage{newtxtext,newtxmath}
\usepackage[capitalise]{cleveref}
\usepackage{tabularx}
\usepackage{caption}
\usepackage{microtype}

\graphicspath{{figs/}}

\crefformat{footnote}{#2\footnotemark[#1]#3}
\begin{document}

\maketitle

\begin{abstract}
Music performance synthesis aims to synthesize a musical score into a natural performance. In this paper, we borrow recent advances in text-to-speech synthesis and present the Deep Performer---a novel system for score-to-audio music performance synthesis. Unlike speech, music often contains polyphony and long notes. Hence, we propose two new techniques for handling polyphonic inputs and providing a fine-grained conditioning in a transformer encoder-decoder model. To train our proposed system, we present a new violin dataset consisting of paired recordings and scores along with estimated alignments between them. We show that our proposed model can synthesize music with clear polyphony and harmonic structures. In a listening test, we achieve competitive quality against the baseline model, a conditional generative audio model, in terms of pitch accuracy, timbre and noise level. Moreover, our proposed model significantly outperforms the baseline on an existing piano dataset in overall quality.
\end{abstract}

\begin{keywords}%
Audio synthesis, computer music, music information retrieval, machine learning, neural network
\end{keywords}

\section{Introduction}
\label{sec:intro}

\begin{figure}
    \centering
    \includegraphics[width=\linewidth]{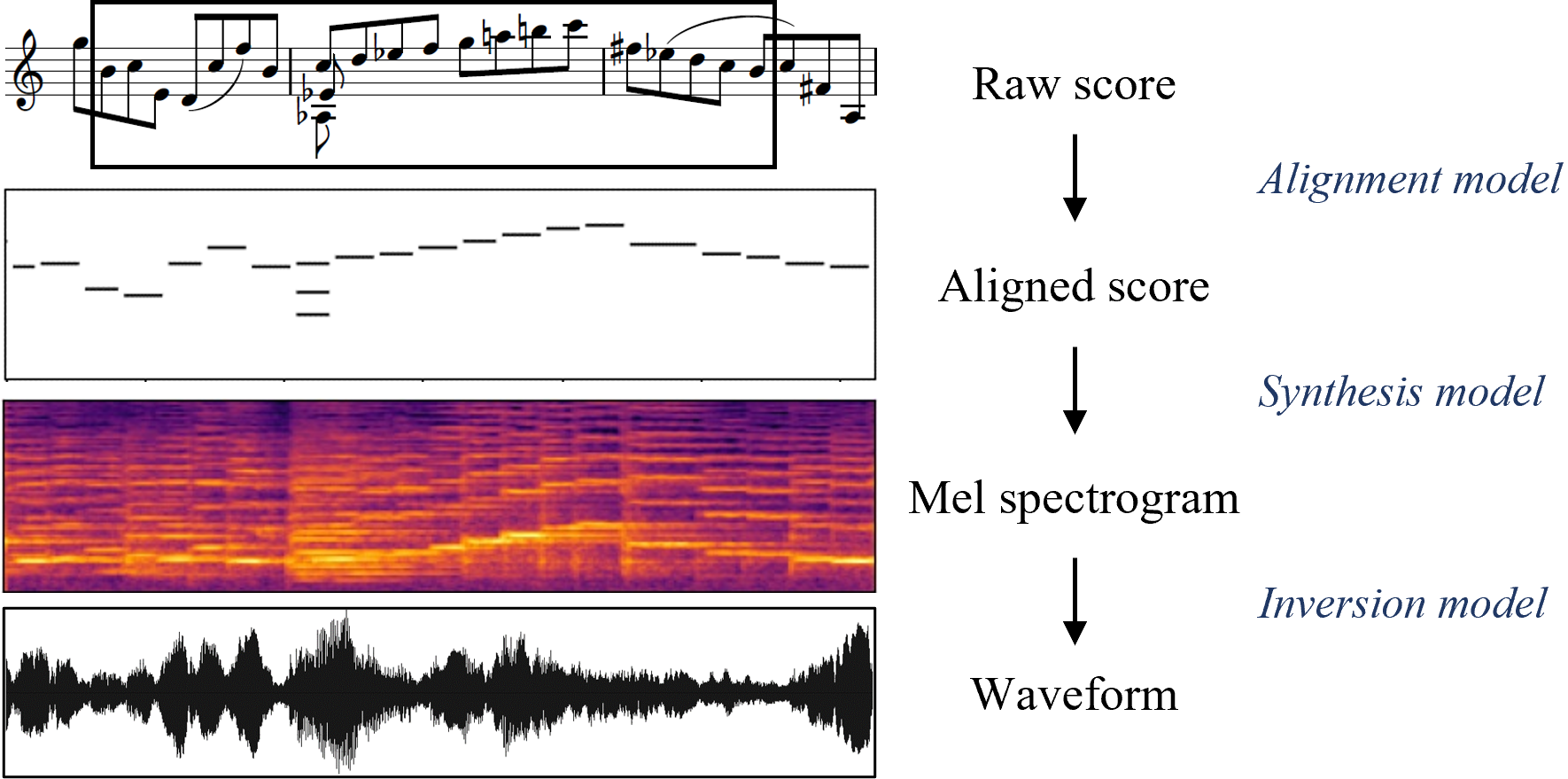}
    \caption{An overview of the proposed three-stage pipeline for score-to-audio music performance synthesis.}
    \label{fig:overview}
\end{figure}

Music synthesis is a complex process that involves both the physics behind a musical instrument and the art of music performance. It remains challenging for a machine to synthesize a natural performance for several reasons. First, it requires a computational model for interpreting and phrasing a musical score. Second, it requires either an explicit or implicit model of the physics and acoustics by which a musical instrument sounds. Third, it requires an understanding of different playing techniques and styles for a musical instrument. While most existing systems only address one of these three challenges at a time, we aim to tackle all these challenges with a data-driven approach using machine learning in this work. We present the Deep Performer---a novel three-stage system for score-to-audio music synthesis, as illustrated in \cref{fig:overview}. 

Prior work has studied music synthesis via various approaches. One line of research focuses on generating realistic samples of musical notes~\cite{engel2017nsynth,defossez2018sing,engel2019gansynth}, while in this work we aim to generate the full performance. Some approach music synthesis by conditioning generative audio models with aligned piano rolls~\cite{manzelli2018wavenet,hawthorne2019maestro}, which we will include as the baseline model in our experiments. Others study synthesizing audio from the fundamental frequency (F0) contour and loudness curve extracted from a recording~\cite{engel2020ddsp,hayes2021waveshaping}, or from lyrics and demo singing audio~\cite{ren2020deepsinger}. On the other hand, some use neural networks to generate expressive timing and dynamics from raw scores~\cite{oore2020performancernn}. Many have also studied inverting mel spectrograms back to waveforms~\cite{shen2018tacotron2,prenger2019waveglow,kumar2019melgan,kong2020hifigan}, including Hifi-GAN~\cite{kong2020hifigan}, which we will use as the inversion model in our proposed system. To the best of our knowledge, prior work on deep neural network based music synthesis either requires an input with expressive timing~\cite{manzelli2018wavenet,wang2019performancenet,hawthorne2019maestro,schimbinschi2019synthnet,kim2019mel2mel,ren2020deepsinger,engel2020ddsp,hayes2021waveshaping} or allows only monophonic (i.e., one pitch at a time) inputs~\cite{ren2020deepsinger,engel2020ddsp,anonymous2022mididdsp}. Our proposed system represents the first that allows unaligned, polyphonic scores as inputs.

In light of the similarity between text-to-speech (TTS) and score-to-audio synthesis, we borrow recent advances from TTS synthesis~\cite{tan2021ttssurvey} to music synthesis and propose a three-stage system for score-to-audio music synthesis. Despite the similarity, music synthesis differs from speech synthesis in that music often contains polyphony, and that long notes are common in music. In order to handle polyphonic music, we propose a new \textit{polyphonic mixer} for aligning the encoder and decoder in a transformer encoder-decoder network~~\cite{vaswani2017transformer,wang2017tacotron}. To provide a fine-grained conditioning to the model, we propose a new \textit{note-wise positional encoding} so that the model can learn to behave differently at the beginning, middle and end of a note. Due to the lack of a proper dataset for training a score-to-audio music synthesis model, we collect and release a new dataset of 6.5 hours of high-quality violin recordings along with their scores and estimated alignments. Through our experiments, we show the effectiveness of our proposed system both qualitatively and quantitatively. Finally, we conduct a subjective listening test to compare our proposed model against a baseline model that uses Hifi-GAN~\cite{kong2020hifigan} to synthesize the waveform directly from an aligned piano roll. Audio samples can be found on our project website.\footnote{\url{https://salu133445.github.io/deepperformer/}\label{fn:website}}

\section{Methods}
\label{sec:methods}

We illustrate in \cref{fig:overview} the proposed three-stage system for score-to-audio music synthesis, which consists of the following three components: (1) an \textit{alignment model} that predicts the expressive timing for each note from a musical score, (2) a \textit{synthesis model} that synthesizes the mel spectrogram from the aligned score, and (3) an \textit{inversion model} that generates the audio waveform given the synthesized mel spectrogram.

\subsection{Alignment model}
\label{sec:alignment-model}

The alignment model consists of a transformer encoder that takes as inputs a sequence of notes and the tempo, followed by a fully-connected layer that outputs the onset and duration of each note. The input score uses metric time with a musically-meaningful unit, e.g., quarter notes, while the output alignment is in the unit of frames. Each note is specified by its pitch, onset, duration and (optional) velocity. In addition, we provide the performer IDs so that the model can learn the different playing styles of performers. The alignment model is trained to minimize the mean squared error (MSE) between the ground truth and predicted onsets and durations, in frames.

\subsection{Synthesis model}
\label{sec:synthesis-model}

\begin{figure}
    \centering
    \includegraphics[width=\linewidth]{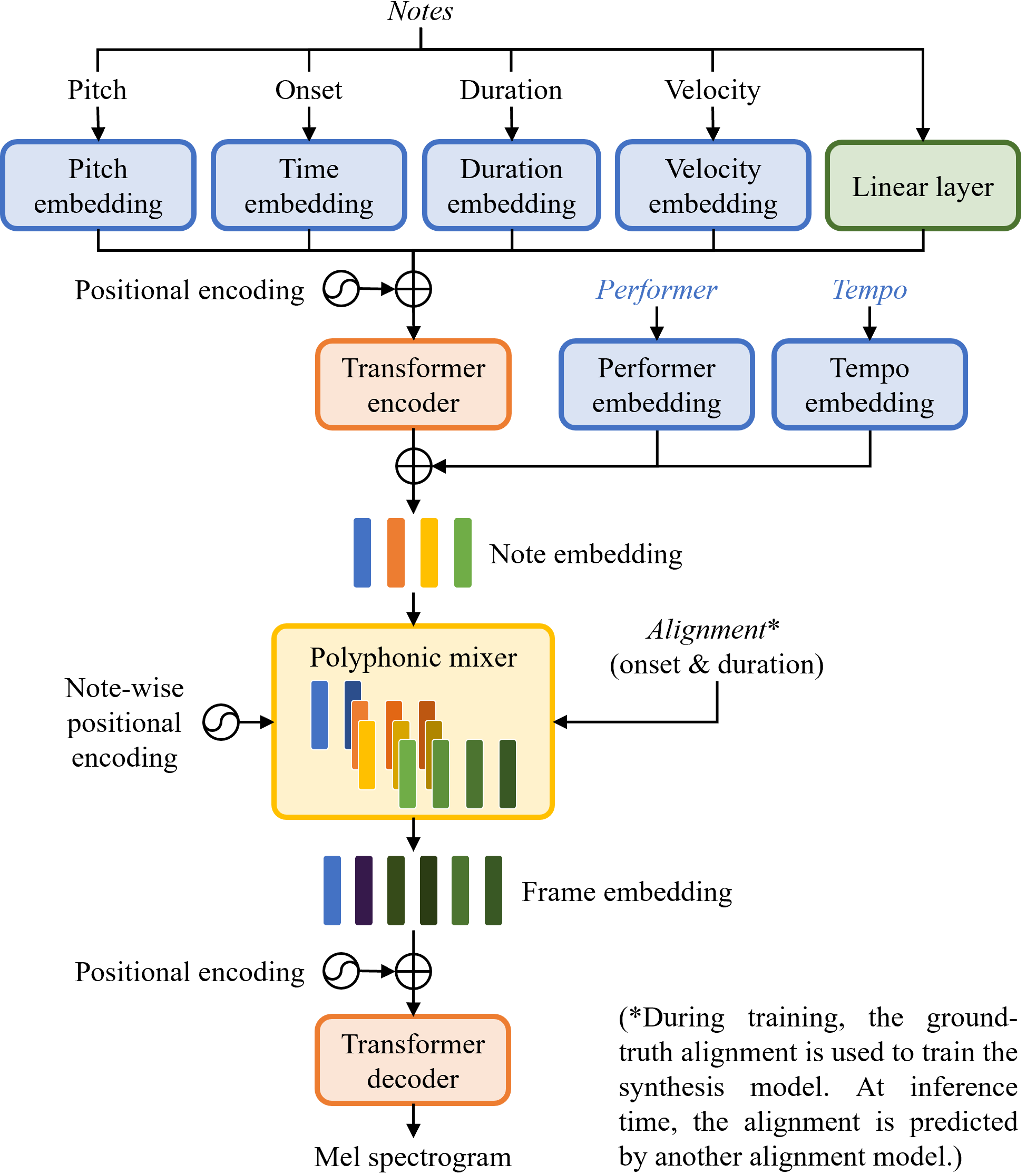}
    \caption{An illustration of the proposed synthesis model.}
    \label{fig:model}
\end{figure}

Given the similarity between TTS and score-to-audio synthesis, we propose a transformer encoder-decoder model for our synthesis model based on~\cite{ren2019fastspeech}. In~\cite{ren2019fastspeech}, each text embedding produced by the encoder is expanded multiple times according to its duration, and then the expanded text embeddings are concatenated to obtain the frame embeddings to be fed to the decoder. This is called the state expansion mechanism~\cite{ren2019fastspeech,yu2020durian}. However, unlike speech, music often contains polyphony. In order to handle polyphonic inputs, we propose the \textit{polyphonic mixer}. As illustrated in \cref{fig:model}, the encoder first encodes the input notes into a sequence of note embeddings. Then, the polyphonic mixer mixes the note embeddings into a sequence of frame embeddings by summing up the note embeddings for the same frame according to their onsets and durations. Finally, the decoder decodes the frame embeddings into a sequence of mel spectrogram frames.

In the state expansion mechanism~\cite{ren2019fastspeech,yu2020durian}, the output vectors remain constant for the duration of a note, and the positional information within each note is missing. However, we argue that such note-wise positional information is critical for the model to behave differently at the beginning, middle and end of a note. Hence, we propose the \textit{note-wise positional encoding} to provide a fine-grained conditioning to the decoder. Mathematically, let $p \in [0, 1]$ be the relative position within a note. For a note embedding $\bm{\mathrm{v}}_{\text{note}}$, we have the corresponding frame embedding at position $p$ as $\bm{\mathrm{v}}_{\text{frame}} = (1 + p \bm{\mathrm{w}}) \odot \bm{\mathrm{v}}_{\text{note}}$, where $\bm{\mathrm{w}}$ is a learnable vector initialized to small random numbers so that $\bm{\mathrm{v}}_{\text{frame}} \approx \bm{\mathrm{v}}_{\text{note}}$ initially. The synthesis model is trained to minimize the MSE between the synthesized mel spectrograms and the ground truth, in log scale.

\subsection{Inversion model}
\label{sec:inversion-model}

Prior work has studied various approaches for synthesizing waveforms from mel spectrograms~\cite{shen2018tacotron2,prenger2019waveglow,kumar2019melgan,kong2020hifigan}. In this work, we adopt the state-of-the-art Hifi-GAN model~\cite{kong2020hifigan} as our inversion model. We note that the proposed three-stage pipeline allows us to use different datasets for training the models. For example, training the inversion model does not require aligned data and thus it can be trained on a larger dataset as unaligned data are relatively easier to acquire.

\section{Data}
\label{sec:data}

\begin{figure}
    \centering
    \begin{minipage}[t]{0.58\linewidth}
        \vspace{0pt}
        \includegraphics[width=\linewidth,height=1.25in]{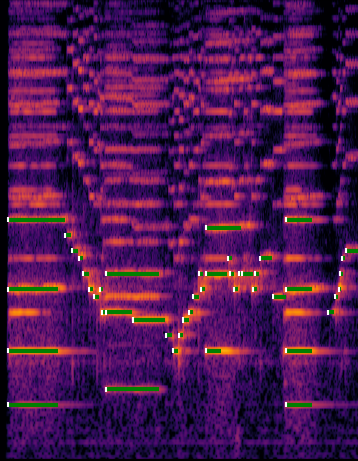}
    \end{minipage}
    \hfill
    \begin{minipage}[t]{0.38\linewidth}
        \caption{An example of the constant-Q spectrogram of the first 20 seconds of a violin recording and the estimated onsets (white dots) and durations (green lines).}
        \label{fig:alignment}
    \end{minipage}
\end{figure}

Due to the lack of a dataset that provides paired audios and scores with fine alignments for training our proposed system, we compile a new dataset of 6.5 hours of professional violin recordings along with their scores and estimated alignments. For copyright concern, we choose Bach's sonatas and partitas for solo violin (BWV 1001--1006) for the ease to acquire high-quality public recordings from the web. The dataset consists of performances by 17 violinists recorded in various recording setups. To acquire the alignment between a recording and its score, we synthesize the scores using FluidSynth~\cite{fluidsynth}, an open-source software synthesizer, with MuseScore General SoundFont~\cite{musescore-soundfont} and perform dynamic time warping on the constant-Q spectrogram of the synthesized audio and that of the recording. We present in \cref{fig:alignment} an example of the dataset and its estimated alignment. To facilitate future research on score-to-audio music synthesis, we release the dataset and the source code for the alignment process to the public.\footnote{\url{https://salu133445.github.io/bach-violin-dataset/}} As discussed in \cref{sec:inversion-model}, the inversion model does not require aligned data for training, and thus we also collect an internal dataset of 156 hours of commercial recordings to train the inversion model. Apart from violin, we also consider the MAESTRO dataset~\cite{hawthorne2019maestro}, which contains 200 hours of piano recordings with finely-aligned MIDI recordings for 10 competition years of the International Piano-e-Competition~\cite{piano-e-competition}. However, since it does not provide the raw scores, we can only train the synthesis and inversion models on this dataset.

\section{Experiments \& Results}
\label{sec:exp}

\subsection{Implementation details}

We use 3 transformer layers in the encoder for the alignment model. The synthesis model shares the same encoder architecture as the alignment model and has 6 transformer layers in the decoder. We use 128 dimensions for all embeddings. For the inversion model, we use the same network architecture as the Hifi-GAN v2 model in~\cite{kong2020hifigan}. We use velocity information only for the piano dataset as it is only available in this dataset. Since performer information is unavailable for the piano dataset, we use the competition years as the performer IDs. We use a temporal resolution of 24 time steps per quarter note for the scores. We downsample the audios to 16 kHz mono and use a hop size of 256 in spectrogram computation, i.e., a temporal resolution of 16 ms. The audios are sliced into 5-second clips for training, where 10\% of them are reserved for validation purpose. We use the Adam optimizer~\cite{kingma2015adam} with a batch size of 16. Unlike \cite{ren2019fastspeech}, we train the alignment and synthesis models separately as we find that joint training hinders convergence. We train the alignment model for 10K steps and all the synthesis models for 100K (violin) and 250K (piano) steps. For each dataset, the inversion model is trained for 1M steps and shared by different synthesis models. We base our implementation on the code kindly released in~\cite{chien2021tts,kong2020hifigan}. We use pretty\_midi~\cite{raffel2014prettymidi} and MusPy~\cite{dong2020muspy} to process the scores.

\begin{figure}
    \centering
    \small
    \begin{tabularx}{\linewidth}{@{}m{.06\linewidth}@{}X@{}}
        (a) &\includegraphics[width=\linewidth,height=.16in]{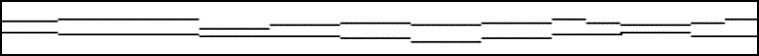}\\[-.5ex]
        (b) &\includegraphics[width=\linewidth,height=.16in]{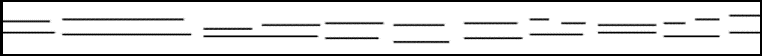}\\[-.5ex]
        (c) &\includegraphics[width=\linewidth,height=.32in]{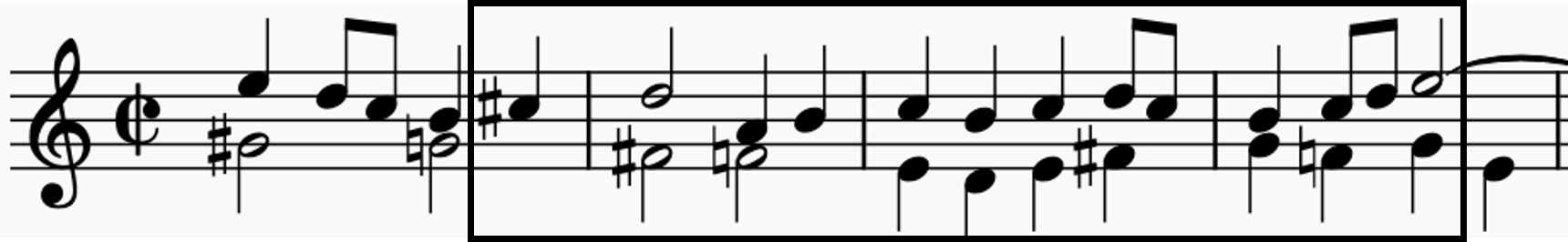}
    \end{tabularx}
    \caption{Examples of the alignments predicted by (a) the constant-tempo baseline model and (b) Deep Performer, our proposed model. (c) shows the input score.}
    \label{fig:results-alignment}
\end{figure}

\begin{figure}
    \small
    \centering
    \begin{tabularx}{\linewidth}{@{}m{.06\linewidth}@{}X@{}}
        (a) &\includegraphics[width=\linewidth,height=.45in]{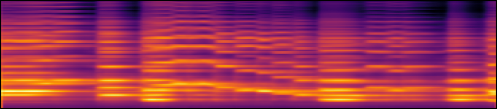}\\[-.75ex]
        (b) &\includegraphics[width=\linewidth,height=.2in]{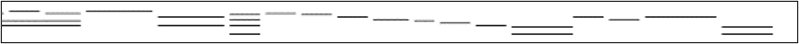}\\[-.25ex]
        (c) &\includegraphics[width=\linewidth,height=.45in]{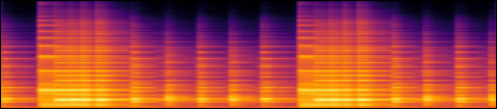}\\[-.75ex]
        (d) &\includegraphics[width=\linewidth,height=.2in]{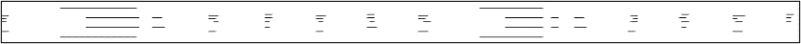}
    \end{tabularx}
    \caption{Examples of the mel spectrograms, in log scale, synthesized by our proposed model for (a) violin and (c) piano. (b) and (d) show the input scores for (a) and (c), respectively.}
    \label{fig:results-synth-poly}
\end{figure}

\begin{table*}
    \small
    \centering
    \caption{Results of the subjective listening test. The mean opinion scores (MOS) and 95\% confidence intervals are reported.}
    \label{tab:results}
    \begin{tabular}{llllll}
        \toprule
                                                          &Violin &&& &Piano\\
        \cmidrule(lr){2-5} \cmidrule(lr){6-6}
                                                          &Pitch accuracy  &Timbre          &Noise level     &Overall         &Overall\\
        \midrule
        Hifi-GAN baseline                                 &4.02 $\pm$ 0.31 &3.13 $\pm$ 0.26 &2.51 $\pm$ 0.29 &2.57 $\pm$ 0.22 &1.49 $\pm$ 0.17\\
        Deep Performer (ours)                             &4.22 $\pm$ 0.30 &3.26 $\pm$ 0.30 &2.67 $\pm$ 0.31 &2.58 $\pm$ 0.21 &2.17 $\pm$ 0.24\\
        ~~~~-~without note-wise positional encoding       &4.13 $\pm$ 0.29 &3.24 $\pm$ 0.27 &2.52 $\pm$ 0.29 &2.61 $\pm$ 0.23 &2.37 $\pm$ 0.23\\
        ~~~~-~without performer embedding                 &3.05 $\pm$ 0.52 &2.54 $\pm$ 0.42 &2.04 $\pm$ 0.31 &2.01 $\pm$ 0.25 &2.26 $\pm$ 0.25\\
        ~~~~-~without encoder (using piano roll input)    &4.30 $\pm$ 0.36 &2.91 $\pm$ 0.28 &2.39 $\pm$ 0.28 &2.22 $\pm$ 0.18 &1.43 $\pm$ 0.16\\
        \bottomrule
    \end{tabular}
\end{table*}

\subsection{Qualitative and quantitative results}
\label{sec:results}

\begin{table}
    \small
    \centering
    \caption{Comparisons of the final MSE between the synthesized mel spectrograms and the ground truths, in log scales.}
    \label{tab:results-loss}
    \begin{tabular}{lll}
        \toprule
                                                          &Violin &Piano\\
        \midrule
        Hifi-GAN baseline                                 &0.892  &0.722\\
        Deep Performer (ours)                             &0.700  &0.436\\
        ~~~~-~without note-wise positional encoding       &0.700  &0.433\\
        ~~~~-~without performer embedding                 &1.030  &0.523\\
        ~~~~-~without encoder (using piano roll input)    &0.844  &0.621\\
        \bottomrule
    \end{tabular}
\end{table}

We show in \cref{fig:results-alignment} an example of the alignment predicted by our proposed alignment model alongside that generated by assuming a constant tempo. We can see that our proposed model is able to predict realistic timing and insert rests between notes. To showcase the effectiveness of the proposed polyphonic mixer, we present in \cref{fig:results-synth-poly} examples of the synthesized mel spectrograms for two polyphonic scores, where we can observe clear harmonic structures and polyphony.

\begin{figure}
    \centering
    \small
    \begin{tabularx}{\linewidth}{@{}m{.06\linewidth}@{}X@{}}
        (a) &\includegraphics[width=\linewidth,height=.5in]{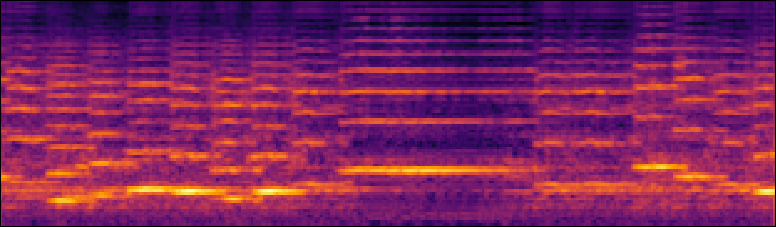}\\[-.25ex]
        (b) &\includegraphics[width=\linewidth,height=.5in]{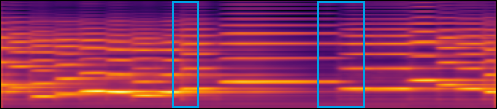}\\[-.75ex]
        (c) &\includegraphics[width=\linewidth,height=.25in]{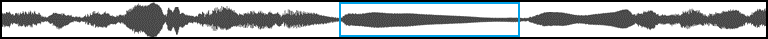}\\[-.25ex]
        (d) &\includegraphics[width=\linewidth,height=.5in]{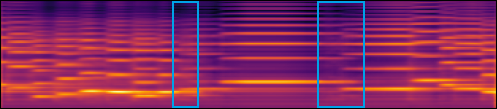}\\[-.75ex]
        (e) &\includegraphics[width=\linewidth,height=.25in]{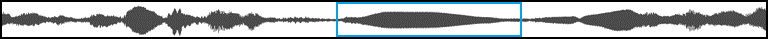}\\[-.25ex]
        (f) &\includegraphics[width=\linewidth,height=.2in]{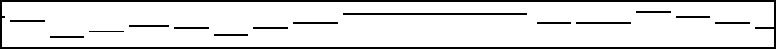}
    \end{tabularx}
    \caption{Examples of the mel spectrograms, in log scale, synthesized by (a) the baseline model, (b) our proposed synthesis model, and (d) our proposed synthesis model without the note-wise positional encoding. (c) and (e) show the waveforms for (b) and (d), respectively. (f) shows the input score.}
    \label{fig:results-comparison}
\end{figure}

Next, we compare our proposed synthesis model against a baseline model that uses a Hifi-GAN~\cite{kong2020hifigan} to synthesize the waveform directly from an aligned piano roll. For a fair comparison, we condition this model with the performer IDs and provide a \textit{position roll} that encodes the note-wise position information. (A position roll is similar to a piano roll, but the values decrease linearly from $1$ to $0$, from the beginning of a note to its end.) As can be seen from \cref{fig:results-comparison}(a) and (b), our proposed model produces smoother contours and clearer harmonic structures, especially on the high frequency end, while the baseline model generates sharper yet noisier results. \cref{tab:results-loss} shows the final MSE between the synthesized mel spectrograms and the ground truths. We can see that our proposed model achieves a lower MSE than the baseline model on both datasets. Finally, due to the reduced temporal resolution of a mel spectrogram compared to that of a waveform, our proposed model is faster in training than the baseline model. Audio samples can be found on our project website.\cref{fn:website}

\subsection{Subjective listening test}

To further evaluate our proposed system, we conduct a subjective listening test with 15 participants recruited from our social networks, where 14 of them plays a musical instrument. We randomly choose 5 musical scores from each dataset and synthesize them with different models. The participants are instructed to rate the synthesized audios in a 5-point Likert scale in terms of pitch accuracy, timbre and noise level as well as the overall quality. We report the results in \cref{tab:results}. We can see that our proposed model significantly outperforms the baseline model on the piano dataset and achieves comparable performance to the baseline on the violin dataset. 

\subsection{Ablation study}

To measure the contributions of different components of the proposed model, we consider three ablated versions of our model. The first removes the note-wise positional encoding. The second removes the performer embedding. The third removes the encoder and uses piano rolls and position rolls (see~\cref{sec:results}) as the inputs to the decoder, while keeping the performer embedding. As we can see from \cref{fig:results-comparison}(b)--(e), note-wise positional encoding help the model produce clearer note transitions and a more realistic waveform envelope (see the highlighted regions). We also report in \cref{tab:results-loss,tab:results} the results for these ablated models. We can see that the performer embedding significantly improves the quality across all criteria. While we show above the effectiveness of the note-wise positional encoding, its impact does not reach statistical significance in our subjective listening test, possibly overshadowed by the artifacts produced by the models. Finally, including an encoder network improves the quality significantly, suggesting that the encoder can learn a more effective representation of the score as compared to the piano roll representation.

\section{Conclusion}
\label{sec:conclusion}

We presented a novel three-stage system for sythesizing natural music performance from unaligned musical scores. We proposed the polyphonic mixer for aligning the encoder and decoder with polyphonic inputs. In addition, we also proposed the note-wise positional encoding for providing a fined-grained conditioning to the synthesis model. Through the subjective listening test, we show that our proposed model significantly outperforms the baseline model on the piano dataset and achieves competitive quality against the baseline on the violin dataset. For future work, we plan to utilize the articulation marks and ornaments on scores to better model playing techniques~\cite{yang2016violin,shih2017violin}, disentangle the timbre from room acoustics to enhance controllability~\cite{engel2020ddsp}, and incorporate adversarial losses~\cite{isola2017pix2pix,yang2021ganspeech} to improve the sharpness of the results.

\section{Acknowledgments}

The authors would like to thank Joan Serrà, Jordi Pons and Lars Villemoes for helpful reviews and discussions.


\clearpage
{
\small
\bibliographystyle{IEEEbib-abbrev}
\bibliography{ref}
}

\clearpage
\appendix

\section{Preprocessing details}

We downmix the recordings to mono and downsample them to 16 kHz using FFmpeg. We then convert them into mel spectrograms using librosa. For the mel spectrogram computation, we use a filter length of 1024, a hop length of 256 and a window size of 1024 in the short-time Fourier transform (STFT), and we use 80 mel bands in mel scale conversion. We summarize these parameters in \cref{tab:preprocessing}.

\begin{table}[h]
    \centering
    \caption{Preprocessing parameters}
    \begin{tabular}{ll}
        \toprule
        Parameter &Value\\
        \midrule
        Audio channels      &mono\\
        Sampling rate       &16 kHz\\
        STFT filter length  &1024\\
        STFT hop length     &256\\
        STFT window size    &1024\\
        Mel bands           &80\\
        \bottomrule
    \end{tabular}
    \label{tab:preprocessing}
\end{table}

\section{Network architectures}

\subsection{Alignment model}

\begin{figure}[ht]
    \centering
    \includegraphics[width=\linewidth]{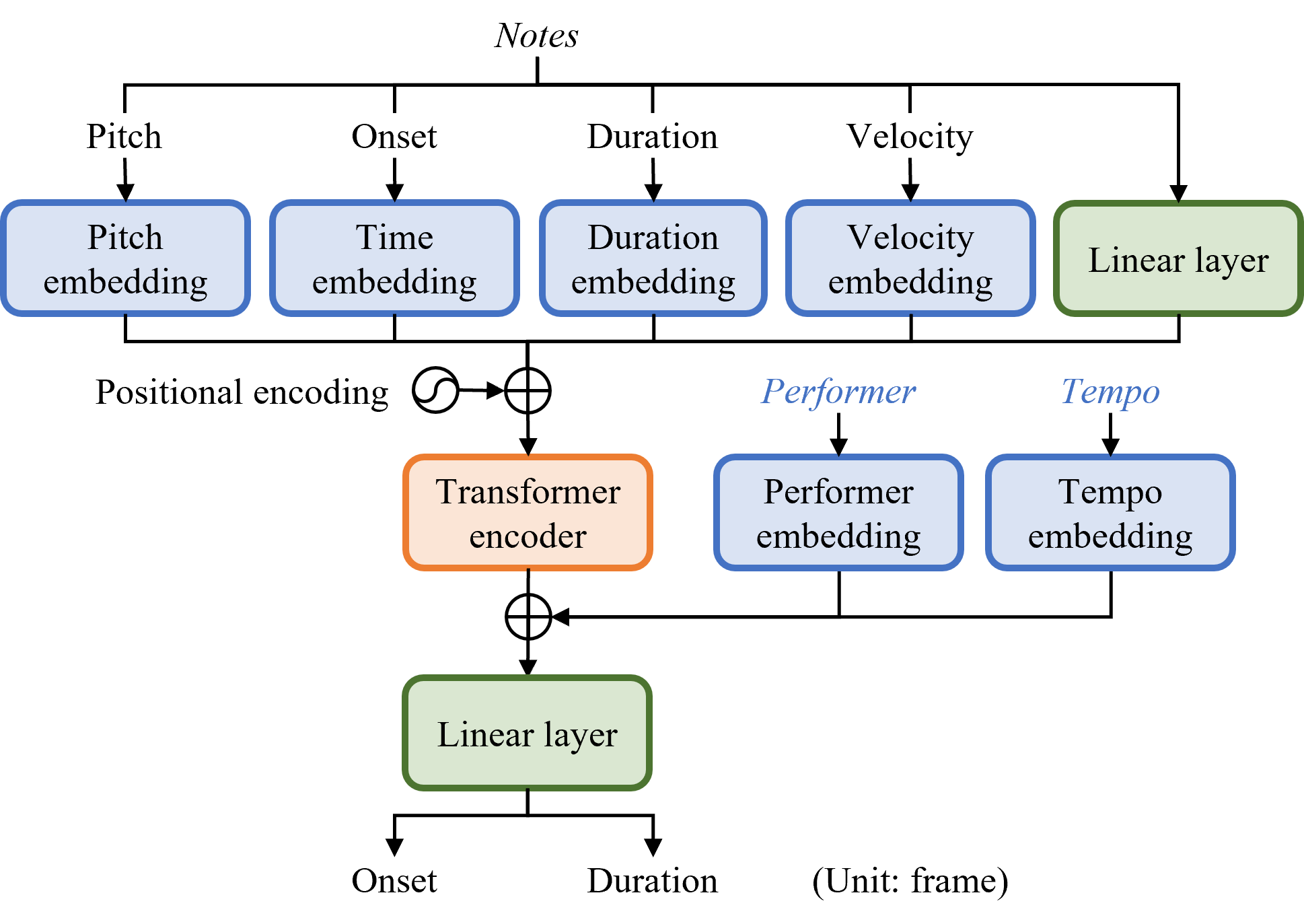}
    \caption{An illustration of the proposed alignment model.}
    \label{fig:alignment-model}
\end{figure}

We illustrate the proposed alignment model in \cref{fig:alignment-model}. We use 128 dimensions for all embeddings. For the transformer encoder, we use 3 transformer layers, each consisting of a multi-head attention (MHA) and a position-wise feed-forward network (FFN) sub-layer. We use 64 hidden neurons and 2 attention heads for each MHA layer. For each FFN layer, we use 256 hidden neurons with kernel sizes of 9 and 1 for the two convolutional layers. Further, we use a maximum sequence length of 1000 and clip the time and duration to 96. We summarize these hyperparameters in \cref{tab:alignment}.

\begin{table}[ht]
    \centering
    \caption{Alignment model architecture}
    \begin{tabular}{ll}
        \toprule
        Parameter &Value\\
        \midrule
        Encoder layers      &3\\
        MHA heads           &2\\
        MHA hidden neurons  &64\\
        FFN hidden neurons  &256\\
        FFN kernel sizes    &9, 1\\
        Max sequence length &1000\\
        Max time            &96\\
        Max duration        &96\\
        \bottomrule
    \end{tabular}
    \label{tab:alignment}
\end{table}

\subsection{Synthesis model}

For the synthesis model, we use 128 dimensions for all embeddings. For the transformer model, we use 3 and 6 transformer layers for the encoder and decoder, respectively. We use 128 hidden neurons and 2 attention heads for each MHA layer. For each FFN layer, we use 256 hidden neurons with kernel sizes of 9 and 1 for the two convolutional layers. In addition, we use a maximum sequence length of 1000. We also clip the time and duration to 96 and 100 for the violin and piano datasets, respectively. We summarize these hyperparameters in \cref{tab:synthesis}. We base our implementation on the source code kindly provided in \cite{chien2021tts}.\footnote{\url{https://github.com/ming024/FastSpeech2}}

\begin{table}[ht]
    \centering
    \caption{Synthesis model architecture}
    \begin{tabular}{ll}
        \toprule
        Parameter &Value\\
        \midrule
        Encoder layers      &3\\
        Decoder layers      &6\\
        MHA heads           &2\\
        MHA hidden neurons  &128\\
        FFN hidden neurons  &512\\
        FFN kernel sizes    &9, 1\\
        Max sequence length &1000\\
        Max time            &96$^*$\\
        Max duration        &96$^*$\\
        \bottomrule
        \footnotesize $^*$100 for the piano dataset
    \end{tabular}
    \label{tab:synthesis}
\end{table}

\subsection{Inversion model}

For the inversion model, we use the network architecture of the Hifi-GAN v2 model proposed in~\cite{kong2020hifigan}. We base our implementation on the source code kindly provided in~\cite{kong2020hifigan}.\footnote{\url{https://github.com/jik876/hifi-gan}}

\subsection{Baseline model}

We base the baseline model on the same Hifi-GAN v2 model~\cite{kong2020hifigan}. In addition, we include an additional linear layer that maps the input piano roll to a hidden vector whose dimension matches the input dimension of the Hifi-GAN v2 model. Further, we include an additional embedding layer to condition the baseline model on the input performer IDs. The outputs of these two layers are summed up and fed as the input to the Hifi-GAN v2 model.

\section{Training details}

We use a batch size of 16 and apply a dropout rate of 0.2 after each sub-layer. We use the same optimizer settings as the original implementation of transformer~\cite{vaswani2017transformer}. For the alignment model, we apply the learning rate annealing schedule used in~\cite{chien2021tts}. We summarize these hyperparameters in \cref{tab:training}. Unlike \cite{ren2019fastspeech}, we train the alignment and synthesis models separately as we find that joint training hinders convergence. For the violin dataset, we train the alignment, synthesis and inversion models for 10K, 100K and 1M steps, respectively. For the piano dataset, we train the synthesis and inversion models for 250K and 1M steps, respectively. For each dataset, the inversion model is trained once and used with different synthesis models.

\begin{table}[h]
    \centering
    \caption{Training hyperparameters}
    \begin{tabular}{ll}
        \toprule
        Parameter &Value\\
        \midrule
        Batch size                          &16\\
        Dropout                             &0.2\\
        Adam optimizer $\beta_1$            &0.9\\
        Adam optimizer $\beta_2$            &0.98\\
        Adam optimizer $\epsilon$           &$10^{-9}$\\
        Gradient clipping threshold         &1.0\\
        Warm up steps (alignment model)     &1000\\
        Warm up steps (synthesis model)     &4000\\
        Learning rate annealing steps$^*$   &10K, 20K, 50K\\
        Learning rate annealing rate$^*$    &0.5\\
        \bottomrule
        \footnotesize $^*$Applied to the alignment model only
    \end{tabular}
    \label{tab:training}
\end{table}

\end{document}